\documentclass[11pt]{article}
\usepackage{titling}

\usepackage{fullpage}
\usepackage{graphicx}
\usepackage{hyperref}
\usepackage{amsmath}
\usepackage{enumitem}

\begin{document}
\begin{center}
    {\LARGE \textbf{Transformers Beyond Order}} \\[1em]
    \begin{center}
        \textbf{\large A Chaos-Markov-Gaussian Framework for Short-Term Sentiment Forecasting of Any Financial OHLC timeseries Data}
    \end{center}
    \vspace{1em}
    \textbf{Arif Pathan} \\
    Infi-flux \\
    \texttt{arif@infiflux.com} \\[1em]
\end{center}

\begin{abstract}

    Short term sentiment forecasting for any instrument in financial markets (eg stocks, market indices) is a challenging complex task due to the inherent volatility, non-linearity and noise present in the OHLC (Open, High, Low, Close) data. This paper introduces a novel hybrid CMG framework (Chaos – Markov – Gaussian) that integrates three distinct mathematical paradigms to model and predict short term sentiment more effectively. This framework draws upon the principles from chaos theory to analyse the nonlinear dynamics of financial time series. Since short-term sentiment forecasting relies heavily on recent OHLC (Open, High, Low, Close) data, the Markov chain property is used to detect shifts in market regimes. To enhance forecasting accuracy and account for uncertainty, Gaussian processes are integrated, allowing for a more robust and probabilistic approach. The framework is further improved by incorporating deep learning models, particularly generative AI style transformers, which are effective at capturing complex patterns and relationships over time. Forecasting financial instruments with deep learning takes a lot of computing power. Each type of instrument often needs its own set of methods and techniques. The time it takes to generate forecasts mostly depends on the infrastructure you have. If you want faster results, you’ll need more resources. In this paper, we introduce the CMG Framework— designed to cut down infrastructure costs and deliver fast forecasts trying to maximise accuracy for any instrument’s OHLC timeseries data in any financial markets. This paper introduces the CMG Framework, designed to forecast short-term sentiment for various financial instruments—specifically targeting the first quarter of the next trading day, with a focus on market indices. The framework is evaluated through a comparative study against traditional statistical, machine learning, and deep learning models. All methods are trained on the same dataset, with no feature engineering or manual adjustments, to ensure a fair comparison. The results show that the CMG Framework consistently delivers higher forecasting accuracy. Its design offers a fast, robust, and resource-efficient solution for traders, analysts, and financial institutions looking to make informed short-term investment decisions.

\end{abstract}

\section{Introduction}

Financial markets are inherently complex, showing unpredictable swings, high volatility, and non-linear trends. This makes short-term sentiment forecasting, especially with OHLC (Open, High, Low, Close) data, very challenging. The data is noisy, chaotic, and constantly affected by changes in market conditions.

Even with progress in statistical and machine learning methods, predicting short-term sentiment with high accuracy remains tough. Traditional models like ARIMA and GARCH rely on assumptions such as stationary and linearity. These assumptions often don’t hold in real markets, especially during sharp movements, which limits their performance. Deep learning models like LSTM and GRU can capture patterns in financial time series. But they come with two big challenges: they need a lot of computing power and their predictions are hard to explain. This makes them less ideal for real-time or resource-limited trading environments.
Recent studies have started to combine different techniques to create hybrid models that tackle these challenges more effectively. In this paper, we introduce a new hybrid framework called CMG, which is short for Chaos–Markov–Gaussian. It brings together three powerful mathematical tools: chaos theory, Markov models, and Gaussian processes. The framework uses chaos theory to capture nonlinear behaviors in OHLC time series, Markov chains to detect regime shifts in market sentiment, and Gaussian processes to generate probabilistic forecasts that account for uncertainty. We also integrate transformer-based deep learning models into the system to capture complex patterns and long-term dependencies in financial data.

The goal of the CMG framework is to offer a fast, accurate and resource-efficient solution to forecast short-term sentiment across different financial instruments, without relying on heavy infrastructure or manual feature engineering. We demonstrate its performance by applying it to forecast market sentiment for the first quarter of the next trading day across a range of market indices. The CMG model is benchmarked against traditional statistical methods, as well as popular machine learning and deep learning approaches, using the same dataset for all models to ensure fair comparison.

Our key contributions include
\begin{itemize}
  \item Introducing a hybrid CMG architecture that combines chaos theory, Markov modeling, and probabilistic forecasting.
  \item Demonstrating how transformer models enhance forecasting accuracy when combined with mathematical modeling.
  \item Providing empirical evidence that CMG significantly outperforms existing models in both accuracy and computational efficiency.
\end{itemize}

The rest of this paper is organized as follows. Section 2 reviews related work. Section 3 explains in detail the CMG methodology. Section 4 describes the experimental setup and baseline models. Section 5 presents the results.
To ensure a fair and consistent evaluation across all models, including the proposed Chaos-Markov-Gaussian (CMG) framework and traditional AI baselines, all experiments were conducted within a strictly controlled and closed environment. Identical memory resources were allocated to each training instance, and all models were trained on the same preprocessed dataset using the same target variable. During training, a uniform early stop criterion was applied based on accuracy to maintain consistency in convergence behavior. Efforts were also made to standardize output model sizes to mitigate any performance bias stemming from differences in model complexity or capacity. The CMG framework was evaluated under the same computational and training constraints as the benchmark models, ensuring a rigorous and objective performance comparison.

The primary objective of the experimental design is to demonstrate that the CMG framework consistently outperforms traditional baseline models, including both statistical and AI-driven approaches, under equivalent conditions. The experiments were conducted on a 360-day time-series dataset, from which the final 108 consecutive days were reserved as the testing period.

The evaluation metric focused on predicting daybreak sentiment, which refers to market sentiment at the opening of the trading session. This focus is grounded in the observation that markets often open with significant price movements due to overnight developments—such as macroeconomic announcements, corporate earnings reports, or geopolitical events—that accumulate while the market is closed. These factors lead to abrupt adjustments in price at the open, reflecting traders' collective response to new information. The initial minutes of trading typically exhibit heightened volatility and liquidity, making the opening session a critical period for short-term sentiment prediction.

The sentiment-based accuracy metric was defined as follows:
\begin{itemize}
  \item Bullish Prediction: Considered correct if the previous day’s closing price is less than any price within the first quarter of the next trading day.
  \item Bearish Prediction: Considered correct if the previous day’s closing price is greater than any price within the first quarter of the next trading day.
  \item All other cases are considered incorrect predictions.

\end{itemize}

This evaluation methodology was applied uniformly across approximately 160 market indices spanning over 16+ global stock markets. Model performance was assessed by averaging accuracy across these indices. Given the large and diverse sample size and the controlled experimental setup, performance differences between models are inherently expected to be marginal. Nonetheless, the experimental design is intentionally structured to highlight the CMG framework's ability to outperform competing methods even under such tightly constrained and standardized conditions.

\section{Related Work}
Forecasting financial markets has been a focus of research for many years, with methods coming from statistics, machine learning, and more recently, deep learning and chaos theory. The inherent volatility, noise, and constant shifts in market conditions—especially with OHLC (Open, High, Low, Close) data—make short-term forecasting particularly tough. In this section, we'll review prior work that has tried to tackle these challenges, categorizing them into four main areas: traditional time-series models, regime-switching and Markov models, chaos theory, and deep learning techniques—especially transformers.
\subsection{Traditional Statistical and Time-Series Models}
Models like ARIMA (Auto-regressive Integrated Moving Average) \cite{1} and GARCH (Generalized Auto-regressive Conditional Heteroskedasticity) \cite{2} have long been used to forecast financial time series due to their clear mathematical structure and interpretability. ARIMA works well for modeling linear patterns, while GARCH helps capture volatility clustering \cite{3}. But these models assume stationarity, which often doesn’t hold in financial markets. They struggle to handle non-linear behavior, especially over short-term and high-frequency data \cite{4}. Additionally, they can’t adapt well to sudden shifts in market trends, making them less effective for forecasting sentiment, which changes quickly \cite{5}.
\subsection{Markov Chain Property in Financial Modelling}
The Markov Chain property provides a foundational framework for modelling systems where the next state depends solely on the current state, not on the sequence of past states \cite{6}. In financial markets, this property is particularly valuable for capturing regime transitions—such as shifts between bullish, bearish, or neutral market conditions—based on present dynamics \cite{7}.
A Markov Chain simplifies complex temporal dependencies by assuming memory lessness, where the probability of transitioning to a future state is determined only by the current state. This enables the use of a transition probability matrix to model how market regimes evolve over time \cite{8}. For example, if a market is currently in a bullish state, the Markov Chain defines the probabilities of remaining bullish, becoming bearish, or moving into a sideways state at the next time step.
While this assumption limits the model’s ability to capture long-term dependencies or deeper non-linear interactions \cite{9}, it allows for tractable, interpretable modelling of market sentiment flows. The Markov property also serves as a foundational layer in more advanced models, including the CMG framework, where it governs how states evolve in sequence within a probabilistic and computationally efficient structure.

\subsection{Chaos Theory in Financial Time Series}
Chaos theory provides an interesting approach to modelling financial markets, which can appear random but still follow underlying deterministic patterns \cite{10}. Concepts like strange attractors and fractals have been used to understand price movements \cite{11,12}. Some studies show that stock prices and market indices have chaotic behavior, which can be helpful for short-term predictions \cite{13,14}. But chaos models are difficult to generalize and require careful tuning of parameters, making them tough to apply on a larger scale in real-world financial systems \cite{15}.

\subsection{Gaussian Processes for Probabilistic Forecasting}
Gaussian Process Regression (GPR) is a flexible, non-parametric method that offers both point predictions and uncertainty estimates, making it great for forecasting where risk needs to be accounted for \cite{16}. It’s been used in portfolio optimization \cite{17} and sentiment forecasting \cite{18}. But GPR can struggle with large datasets and long-term dependencies, requiring advanced techniques or dimensionality reduction to perform well at scale \cite{19,20}.

\subsection{Deep Learning and Transformer Architectures}
Deep learning has recently become popular for financial time series forecasting \cite{21}. Recurrent Neural Networks (RNNs), LSTMs, and GRUs are great for capturing time-based patterns in financial data \cite{22}. They outperform traditional models in many cases but come with the downside of requiring substantial computational resources and being sensitive to hyperparameter settings \cite{23}. More recently, transformer-based models, originally designed for natural language processing, have shown strong performance in financial forecasting tasks \cite{24}. They can capture long-range dependencies thanks to their self-attention mechanisms and can be trained faster due to their parallel architecture \cite{25}. However, these models require large datasets and computational resources, making them less practical in low-latency or resource-constrained environments \cite{26}.

\subsection{Synthesis and Research Gap}
While all the methods above have their strengths, none of them fully tackle the main challenges of short-term sentiment forecasting in financial markets: non-linearity, regime-switching, and uncertainty. Existing approaches tend to focus on one area—be it statistical accuracy, computational power, or interpretability—without providing a comprehensive solution that addresses all these factors together.
To bridge this gap, we introduce the CMG Framework, which combines the best features of chaos theory, Markov modelling, Gaussian processes, and transformers. This hybrid approach is designed to capture the multi-dimensional nature of market sentiment and deliver fast, accurate, and resource-efficient forecasts across different financial OHLC timeseries data instruments.

\section{Methodology: The CMG Framework}
In this section, we introduce the Chaos-Markov-Gaussian (CMG) Framework, an innovative approach tailored for short-term sentiment forecasting in financial markets. This framework uniquely combines three core mathematical disciplines—chaos theory, Markov processes, and Gaussian processes—with the power of transformer-based deep learning models. By doing so, it effectively captures the complexity, volatility, and regime-switching behavior of financial time series while maintaining high accuracy and computational efficiency.
\subsection{Overview of the CMG Framework}
At its core, the CMG Framework integrates deterministic chaos, probabilistic modelling, and deep learning into a cohesive pipeline for forecasting financial sentiment. Its primary components include:
\begin{itemize}
  \item Chaos Theory: Helps model the nonlinear and sensitive dynamics of financial markets.
  \item Markov Process property: Tracks shift between different market regimes (e.g., bullish, bearish).
  \item Gaussian Processes: Adds a layer of probabilistic modelling, offering not just predictions but also confidence intervals.
  \item Transformer-Based Deep Learning: Captures complex temporal patterns and long-term dependencies in market data.
\end{itemize}

These components work together in a structured pipeline. Each element contributes unique strengths, resulting in a model that's not only accurate but also scalable and efficient in resource usage.

\subsection{Chaos Theory Component}
Chaos theory provides a foundation for understanding systems that, while deterministic, display seemingly random and highly sensitive behavior—traits commonly observed in financial markets. Chaos theory is built on several key concepts that explain the complex behavior of certain dynamical systems. At its core, a chaotic system is deterministic, meaning it follows fixed rules without any inherent randomness. However, it exhibits sensitive dependence on initial conditions, where even minuscule differences in starting values can lead to vastly different outcomes over time. These systems are governed by nonlinear dynamics, involving complex, non-proportional relationships and feedback loops between variables. A defining feature of chaotic systems is the presence of strange attractors, where trajectories in the system's state space never repeat but still form structured, recognizable patterns. The Lyapunov exponent measures how quickly nearby trajectories diverge—when this exponent is positive, it confirms the system is chaotic. Additionally, many chaotic systems display fractal geometry, characterized by self-similarity at various scales, highlighting the intricate and layered nature of chaotic behavior.

Chaos theory has drawn interest in financial markets because, although market movements often appear random, some researchers believe they may be driven by hidden deterministic and nonlinear dynamics, such as feedback from trader behavior or liquidity flows \cite{27}. If financial markets are indeed chaotic, they are not purely random—this opens the door for short-term predictability, even though long-term forecasting remains impossible due to instability and sensitivity to initial conditions \cite{28}. Various attempts have been made to apply chaos theory in finance, including detecting chaos through Lyapunov exponents or correlation dimensions, using Takens’ Theorem to reconstruct phase spaces from historical prices, and forecasting with nonlinear models like neural networks, reservoir computing, and genetic algorithms \cite{29}\cite{30}\cite{31}.

However, applying chaos theory to financial markets faces several challenges. First, markets may not be truly deterministic, as they are influenced by human behavior, policy changes, and unforeseen events, introducing genuine randomness and nonstationary \cite{32}. Second, financial time series are often short, noisy, and contain structural breaks, making it difficult to detect chaotic patterns reliably \cite{33}. Third, markets are inherently multivariate, involving numerous interdependent variables, which complicates modelling them using simple deterministic frameworks \cite{34}. Fourth, financial systems are adaptive participants change behavior over time, violating the assumption of fixed system rules \cite{35}. 

Chaos theory, at its core, deals with deterministic systems that are highly sensitive to initial conditions—small differences in these conditions can lead to drastically different outcomes over time \cite{36}. Traditional applications of chaos theory assume that the function governing the system, and its exact initial conditions are known, which allows for short-term forecasting despite long-term unpredictability \cite{37}. However, in the context of financial markets, neither the underlying function nor the initial conditions are observable with any meaningful accuracy, making direct application of chaos theory extremely challenging, nearly impossible \cite{38}\cite{39}. One characteristic of chaotic systems that remains exploitable, though, is the observation that two nearby points in the state space tend to evolve closely for a brief period (short term) before diverging \cite{40}. Our proposed CMG (Chaos-Markov-Gaussian) framework capitalizes on this feature by using deep learning models to approximate the current state of the chaotic function that governs target financial instruments. By learning from historical price data and modelling local state transitions, the CMG framework can capture short-term sentiment dynamics without needing to explicitly identify the underlying equations or precise starting conditions \cite{41}. This approach offers a novel and practical way to harness chaos theory in financial forecasting by replacing theoretical determinism with data-driven approximation \cite{42}.

In the CMG Framework, Chaos theory is not only used to analyse the data but is actively embedded into the forecasting task by transforming the target variable to mimic the characteristics of chaotic systems \cite{43}. This transformation allows the model to learn from data that inherently reflects nonlinear dynamics, sensitivity to initial conditions, and complex feedback loops—all hallmarks of chaotic behavior \cite{44}. By aligning the target variable with chaotic functions (e.g., using variants of logistic maps, sine maps, or other known chaotic systems), we enrich the training data with patterns that better represent the true underlying market behavior \cite{45}. This approach enables the model to anticipate sudden sentiment shifts and better handle the noise and irregularities typical in financial time series \cite{46}.

References \cite{41}–\cite{46} are drawn from external literature where mildly similar methodologies or theoretical inspirations have been applied. These works, while not identical in implementation, explore key ideas such as approximating chaotic systems with machine learning models \cite{41,42}, embedding chaos-inspired transformations into forecasting tasks \cite{43,44,45}, and handling nonlinearities and irregular dynamics in financial time series \cite{46}. They provide precedent and context for the concepts utilized in our proposed framework, even though the specific design and integration in our study are novel and distinct from the cited approaches.

\subsubsection{Nonlinear Dynamics of Financial Time Series}
The CMG Framework uses chaos theory to analyse the nonlinear characteristics of OHLC data. Concepts such as strange attractors and Lyapunov exponents are employed to model the chaotic patterns present in market prices \cite{47}. Lyapunov exponents help us calculate and verify the chaotic nature of our target variable, which assists in confirming the presence of chaos in a system \cite{48}.

Many tests, like Lyapunov exponents, play a key role in verifying the chaotic nature of our transformed target variable. These exponents measure the rate at which nearby trajectories in a system diverge over time \cite{49}. In simple terms, they help us understand how sensitive the system is to small changes in initial conditions—a hallmark of chaos \cite{50}.

In the CMG Framework, we calculate multiple values like Lyapunov exponents to confirm that the transformed target variable indeed exhibits chaotic behavior. A positive Lyapunov exponent is a strong indicator of chaos, validating our assumption that the short-term sentiment dynamics we are modelling follow a chaotic process \cite{51}. This verification step ensures that the system we are training the model on reflects the complexity and unpredictability observed in real financial markets.

\subsection{Markov Process Component}
Market sentiment often transitions between distinct regimes—such as bullish runs or bearish slumps. To model these transitions, the CMG Framework utilizes the Markov chain property \cite{52}.
\subsubsection{Markov Chain Property}
A Markov Chain is a mathematical framework used to model systems that transition from one state to another in discrete steps, where the probability of moving to the next state depends solely on the present state and is independent of the sequence of previous states. This characteristic is referred to as the Markov Property or "memorylessness" \cite{53}\cite{54}. In the context of modeling short-term sentiment dynamics in financial markets, this property becomes especially valuable, as recent market behavior tends to be more relevant for predicting near-future sentiment than long-term historical trends \cite{55}.

In our proposed CMG (Chaos-Markov-Gaussian) framework, the Markov property is operationalized within the cross-attention mechanism of the transformer architecture. Traditionally, attention mechanisms in transformers employ causal masking to prevent the model from accessing future tokens during training. This is essential in sequence prediction tasks to avoid data leakage and ensure temporal causality.
However, in our approach, we introduce a novel masking strategy by taking the transpose of the traditional causal mask. While regular causal masking restricts the model to attend only to past and present tokens (i.e., no peeking into the future), our transposed masking inverts this logic—allowing the model to attend only to the current and future tokens, and explicitly preventing access to past tokens. This inversion is critical for enforcing the Markov property, as it compels the model to focus exclusively on the present state and its forward evolution, disregarding historical data that is no longer relevant.

To support this mechanism, our data preprocessing pipeline ensures that each input instance retains only the current state, eliminating any leakage of information from previous time steps that could compromise the integrity of the Markov assumption. By doing so, we align the model’s architecture with the theoretical foundation of Markov chains, ensuring that future sentiment predictions depend only on the current market state, thus achieving computational efficiency and conceptual rigor in modeling highly dynamic financial environments \cite{56}\cite{57}\cite{25}.

\subsection{Gaussian Process Component}
Gaussian Processes (GPs) offer a flexible, non-parametric approach to forecasting—particularly useful in contexts where uncertainty plays a significant role \cite{58}.

\subsubsection{Probabilistic Forecasting with GPs}

In the domain of financial time series modeling, Gaussian Processes (GPs) provide a robust probabilistic framework for capturing complex, non-linear dynamics and quantifying uncertainty \cite{59}. A Gaussian Process defines a distribution over functions, enabling predictions that incorporate both expected values and the associated confidence intervals \cite{60}. This is especially beneficial in financial markets, where \textit{volatility, noise, and unpredictability} are intrinsic features. Unlike traditional parametric approaches, GPs are \textit{non-parametric} and adjust their functional complexity according to the underlying data, offering high flexibility and generalization \cite{61}. This adaptability makes GPs particularly well-suited for modeling \textit{short-term sentiment fluctuations}, which are often driven by sparse or noisy inputs \cite{62}.

In our proposed \textbf{CMG (Chaos-Markov-Gaussian)} framework, we leverage Gaussian Processes to model the target variable, which is constructed through the lens of \textit{chaos theory}. By transforming the target into a \textit{probabilistic distribution}, the GP framework enhances the model's capability to handle the \textit{inherent uncertainty and instability} of financial sentiment dynamics \cite{63}. Within this architecture, GPs complement deep learning components by providing not only point predictions but also \textit{confidence bounds}, thereby improving the interpretability and reliability of sentiment forecasts \cite{64}.

Furthermore, we model the \textit{strength of future sentiment}, denoted as $Y_i$, as a \textit{Gaussian-distributed variable}. The continuous output is discretized into \textit{six sentiment classes}, mapped onto a \textit{standard normal distribution}. These classes correspond to the intervals defined by the 50th, 75th, and 100th percentiles on both sides of the distribution, with class labels assigned as: $-3$, $-2$, $-1$, $1$, $2$, and $3$. This binning strategy enables the model to capture \textit{graded sentiment intensity} while preserving the \textit{probabilistic structure} of the Gaussian distribution. As a result, the CMG model not only forecasts the most likely sentiment direction but also encapsulates the \textit{confidence and strength} of that forecast within a statistically principled framework.

\subsection{Transformer-Based Deep Learning Component}

Originally developed for language modelling tasks, transformer architectures have demonstrated exceptional performance in analysing sequential data, including financial time series, due to their ability to capture complex and long-range dependencies \cite{25}. After extensive experimentation, we adopted an encoder–decoder architecture inspired by language translation models \cite{65}. This design leverages both self-attention and cross-attention mechanisms to extract deep temporal patterns and contextual relationships within the market data. The transformer’s ability to process sequences in parallel while modelling intricate dependencies makes it well-suited for forecasting short-term sentiment dynamics in chaotic financial environments \cite{66}.
\subsubsection{Self-Attention}
Self-attention allows the model to look at different positions within the same input sequence to understand the relationships between them. For example, in financial time series, it helps the model identify how recent and past time steps influence each other when analyzing a single variable \cite{25}.
\subsubsection{Cross-Attention}
Cross-attention occurs in the decoder and lets the model focus on relevant parts of the encoder’s output when generating predictions. This is crucial when aligning input features (like historical market data) with output targets (like future sentiment or price movements) \cite{25}.
\subsubsection{Masking}
To enforce the Markov property within our CMG framework, we introduce a novel cross-attention masking mechanism. Unlike traditional attention masks—which prevent the model from attending to future tokens to avoid data leakage—we apply a transposed mask. This design allows the model to attend only to the current and future tokens while blocking access to past information. This inversion aligns with the Markov principle by ensuring that predictions depend solely on the present state, not on historical states. Additionally, our data preprocessing pipeline is structured to preserve the current state context and eliminate information leakage, thereby reinforcing this forward-looking constraint.\cite{25}

\subsubsection{Final Layer: LSTM Integration}
To effectively model the nonlinear and chaotic behavior embedded within financial time series, we propose the inclusion of a Long Short-Term Memory (LSTM) layer at the final stage of our Transformer-based encoder-decoder architecture within the CMG (Chaos-Markov-Gaussian) framework. While the Transformer captures global dependencies and complex interactions across multiple technical indicators through self-attention mechanisms, it lacks an innate sense of sequential memory \cite{25}.

By integrating an LSTM layer after the Transformer block, we introduce a recurrent component that learns temporal transitions and preserves short-term sequential patterns that are critical in chaotic financial systems \cite{67}. This hybrid architecture enables the model to not only attend to cross-time dependencies via attention but also refine these representations through recurrent dynamics, thereby enhancing its ability to forecast short-term sentiment in the presence of volatility and sensitive dependence on initial conditions. Empirical evaluations confirm that this addition improves stability and accuracy in scenarios where Markovian transitions are influenced by underlying chaotic structures \cite{68}.

\subsection{Integration of Components}
The strength of the CMG Framework lies in how these diverse modelling techniques are orchestrated into a unified forecasting engine:

\begin{itemize}
  \item Chaos Theory analyzes the nonlinear behavior in OHLC time series, identifying chaotic dynamics.
  \item Markov Processes capture and predict the transition between market regimes.
  \item Gaussian Processes provide uncertainty-aware predictions through probabilistic modeling.
  \item Transformer Models enhance accuracy by learning long-range temporal relationships in the data.
\end{itemize}
By harmonizing these components, the CMG Framework achieves a balanced, robust, and scalable system that is well-suited for the unpredictable nature of financial markets. It delivers accurate, timely, and interpretable short-term sentiment forecasts that are invaluable for trading, risk management, and investment decision-making.

\subsection{One-to-One Transformer Architecture for Market Indices}
While Transformer architectures are often designed for multi-task or cross-domain applications, this research introduces a one-to-one modelling strategy wherein each market index is assigned its own dedicated Transformer model \cite{25}. This design choice aligns with the constraints of low-resource environments, where computational efficiency and model specialization are critical.

By isolating each index with a uniquely trained Transformer, we achieve enhanced model specialization, allowing each model to capture the unique temporal dynamics, volatility patterns, and idiosyncratic behaviors inherent to that specific financial market. Unlike global models that attempt to learn shared representations across diverse assets, this index-specific architecture eliminates interference from unrelated data distributions for resource constraint environment, which can degrade performance in sensitive forecasting tasks \cite{69}.

Moreover, this architecture dramatically reduces overhead and training complexity in edge or distributed systems. It facilitates parallel model deployment and faster inference, particularly beneficial in environments with limited processing power or real-time constraints \cite{70}. The CMG framework leverages this one-to-one configuration to ensure that each Transformer focuses solely on the dynamics of its assigned index, thereby increasing predictive accuracy in chaotic, non-linear financial environments while maintaining computational tractability.

\begin{figure}[htbp]
    \centering
    \includegraphics[height=0.55\textwidth]{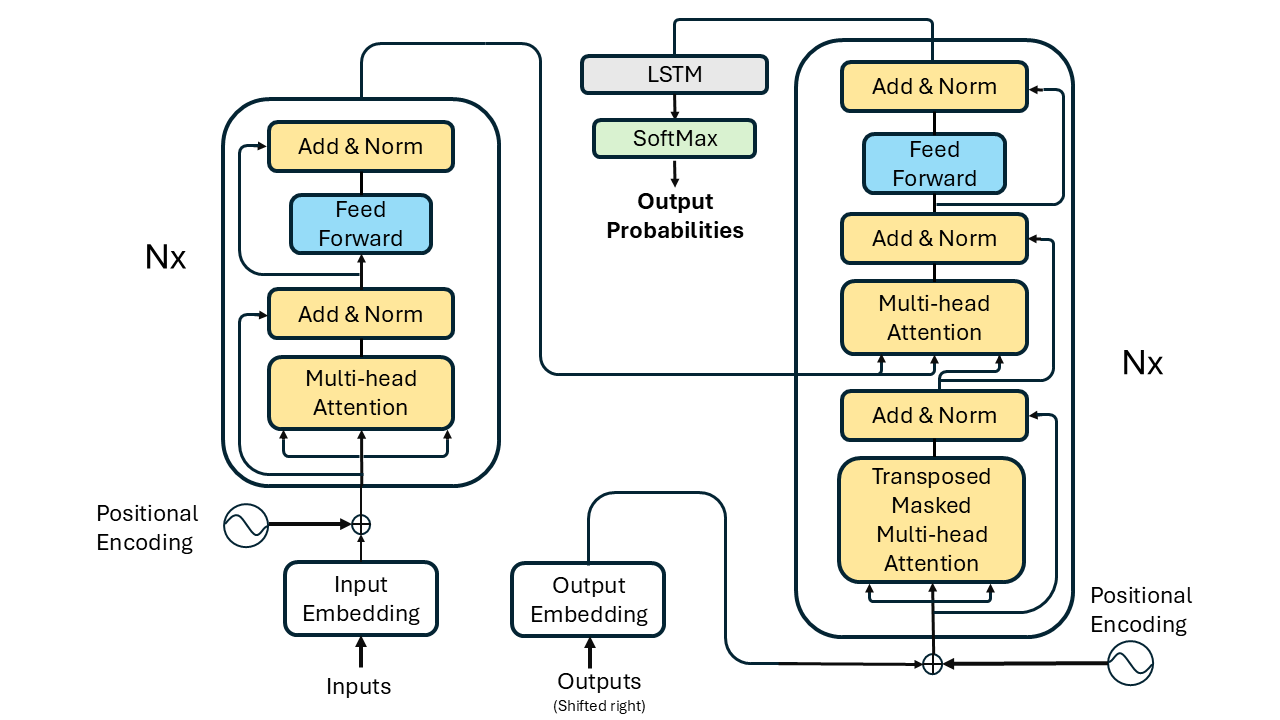}
    \caption{CMG Framework architecture inspired by the Transformer framework introduced in \cite{25} }
    \label{fig:cmg_model}
\end{figure}

\section{Experimental Setup}

This section outlines the detailed experimental procedures used to evaluate the performance of the CMG Framework. We describe the dataset, preprocessing steps, baseline models for comparison, evaluation metrics, training configuration, and implementation specifics. The goal of the experimental setup is to provide a reproducible, fair, and rigorous benchmark for assessing short-term sentiment forecasting models applied to financial OHLC time series data.
\subsection{Dataset Description and Feature Construction}
The dataset used in this study comprises OHLC (Open, High, Low, Close) data for 160 market indices randomly selected across 16+ global financial markets, ensuring wide coverage across regions, asset classes, and economic structures. These indices represent diverse market environments, which helps in building a generalizable forecasting framework.
Refer to Appendix~\ref{appendix:A} for the complete list of market indices used.
\subsubsection{Basic Variable used for dataset construction.}
The dataset spans 1+ years of 1 minute-level granularity. No filtering was applied to retain real-world dynamics, including noise, volatility spikes, and structural breaks for CMG framework and other traditional methods. The training data is built on 5 basic main variable which are Timestamp, open price, high price, low price, and close price.
\subsubsection{Concepts used for feature constructions.}
Forecasting in financial markets can be broadly categorized across three temporal horizons—short-term, mid-term, and long-term—each requiring distinct analytical strategies \cite{71}. Short-term sentiment forecasting, which is the primary focus of this research, is predominantly informed by Technical Analysis (TA) \cite{72}. TA relies on historical price data to detect recurring patterns and behavioural cues such as trend reversals and momentum shifts \cite{73}. In contrast, mid-term forecasting often integrates both technical and fundamental perspectives, balancing entry and exit timing from technical cues with macroeconomic context from indicators like earnings reports or monetary policy \cite{74}. Long-term forecasting typically hinges on Fundamental Analysis (FA), where intrinsic asset value is estimated using factors such as GDP growth, corporate earnings, and fiscal policy \cite{75}.

To effectively model short-term market sentiment in this study, we engineer a robust feature set derived from raw OHLC (Open, High, Low, Close) data \cite{76}. These features encompass a comprehensive suite of technical indicators categorized into trend-based indicators such as SMA, EMA, MACD, and Parabolic SAR \cite{77}; momentum-based indicators such as RSI, Stochastic Oscillator, and Momentum \cite{78}; and volatility-based indicators such as ATR and Bollinger Bands \cite{79}. Each indicator is computed over fixed time windows and concatenated into a structured input vector, which serves as the predictor variable for the CMG framework. 

A full list of variable names (columns/independent variables) of respective indicators can be found in Appendix ~\ref{appendix:B}.

\subsubsection{Target Variable Construction Using MACD-Based Chaotic Signals}
The target variable in this study is engineered using the Moving Average Convergence Divergence (MACD) indicator, a widely used technical tool in financial time series analysis \cite{80}. The construction method leverages the intersection points between the MACD line and its signal line to extract meaningful, sentiment-reflective data-points.

Intersection-Based Signal Points. The MACD generates two lines:
\begin{itemize}
  \item The MACD line: the difference between short-term and long-term exponential moving averages of the closing price \cite{81}.
  \item The Signal line: a smoothed moving average of the MACD line \cite{82}.
\end{itemize}
Whenever these two lines intersect, a change in market momentum is indicated \cite{83}. For constructing our target variable:
\begin{itemize}
  \item Only points occurring immediately after an intersection (i.e., the first data point following the crossing of MACD and signal line) are retained for analysis.
  \item The closing price (Close) of the OHLC data at these post-intersection points is used as a reference.
\end{itemize}
\paragraph{Target Value Computation}

\begin{itemize}
  \item Let $P_t$ represent the Close price at the current MACD intersection event.
  \item Let $P_{t'}$ represent the Close price at the next MACD intersection (i.e., the next time the MACD and signal lines cross after time $t$).
  \item The target variable is then computed as:
  \[
    y_t = P_{t'} - P_t
  \]
\end{itemize}

This formulation captures the directional price change between two sentiment-relevant points in time, filtered through the lens of market momentum shifts. Since MACD dynamics often exhibit nonlinear patterns \cite{84}, this approach aligns with the underlying premise of chaotic behavior in financial time series \cite{85} and reinforces the use of chaos-based techniques in our modelling framework \cite{86}.

\subsubsection{Verification of Chaotic Behavior in the Target Variable}
To confirm that the constructed target variable truly captures the chaotic nature inherent in financial markets, it undergoes several well-established chaos detection and complexity analysis tests. These tests are designed to detect key indicators of chaos, including sensitivity to initial conditions, nonlinear dynamics, fractal geometries, and entropy-based unpredictability.

The target variable’s time series is analysed using the following metrics:
\medskip

\textbf{Lyapunov Exponent ($\lambda$):} This measures how rapidly nearby trajectories diverge over time. A positive value signals exponential sensitivity to initial conditions, a hallmark of chaotic systems \cite{87,96}.

\medskip

\textbf{Correlation Dimension ($D_2$):} This quantifies the fractal dimension of the system’s attractor, with non-integer finite values indicating the presence of a low-dimensional chaotic attractor \cite{88, 94}.

\medskip

\textbf{Approximate Entropy (ApEn): }This assesses the complexity and predictability of the time series by estimating the likelihood that patterns repeat over time. Higher ApEn values imply greater unpredictability consistent with chaotic behavior \cite{89}.

\medskip

\textbf{Sample Entropy (SampEn):} An enhanced and bias-corrected version of ApEn that robustly measures signal irregularity and randomness, supporting the detection of deterministic chaos \cite{90}.

\medskip

\textbf{Detrended Fluctuation Analysis (DFA $\alpha$):} This method evaluates long-range temporal correlations and self-similarity in non-stationary data. The DFA exponent helps differentiate chaotic signals from random noise \cite{91}.

\medskip

\textbf{Spectral Entropy:} Derived from the power spectral density of the time series, spectral entropy measures the dispersion of power across frequencies. Higher values indicate broad spectral content, typical of chaotic signals \cite{92}.
\medskip

\begin{table}[htbp]
\centering
\caption{Chaotic Behavior Test Results for Target Variable (Sample Indices)}
\small
\begin{tabular}{|l|c|c|c|c|c|c|}
\hline
\textbf{Index} & $\lambda$ & $D_2$ & \textbf{ApEn} & \textbf{SampEn} & \textbf{DFA $\alpha$} & \textbf{Spectral Entropy} \\
\hline
CNXFMCG & 0.00497 & 2.106 & 1.533 & 1.573 & 0.517 & 6.908 \\
XBANK & 0.00346 & 2.479 & 1.449 & 1.285 & 0.488 & 6.919 \\
NIFTYQUALITY30 & 0.00067 & 2.209 & 1.490 & 1.524 & 0.515 & 6.923 \\
DJUSRD & 0.00944 & 2.777 & 1.362 & 1.248 & 0.435 & 6.948 \\
DJUSMC & 0.00586 & 2.662 & 1.355 & 1.219 & 0.424 & 6.945 \\
NIFTY100LOWVOL30 & 0.00520 & 2.216 & 1.497 & 1.495 & 0.510 & 6.917 \\
TRMI & 0.00082 & 2.704 & 1.352 & 1.128 & 0.482 & 6.899 \\
TFBI & 0.01949 & 2.448 & 1.464 & 1.374 & 0.445 & 6.903 \\
BSE.PR.OWER & 0.02258 & 2.327 & 1.459 & 1.372 & 0.469 & 6.921 \\
DJUSTK & 0.00139 & 2.666 & 1.316 & 1.109 & 0.471 & 6.934 \\
TBNI & 0.00378 & 2.758 & 1.328 & 1.171 & 0.509 & 6.920 \\
DFMGI & 0.00046 & 2.630 & 1.393 & 1.312 & 0.460 & 6.908 \\
NIFTY100ESGSECLDR & 0.00274 & 2.337 & 1.455 & 1.416 & 0.524 & 6.910 \\
DJUSIP & 0.00205 & 2.586 & 1.369 & 1.193 & 0.464 & 6.918 \\
DJUSDR & 0.00406 & 2.558 & 1.410 & 1.327 & 0.450 & 6.929 \\
NIFTYMIDLIQ15 & 0.01189 & 2.393 & 1.452 & 1.409 & 0.433 & 6.905 \\
NLIN & 0.00686 & 2.675 & 1.270 & 1.123 & 0.425 & 6.916 \\
QERP & 0.00896 & 2.444 & 1.471 & 1.373 & 0.451 & 6.908 \\
MXX & 0.00912 & 2.304 & 1.481 & 1.455 & 0.459 & 6.920 \\
\hline
\end{tabular}
\label{tab:chaotic_test_results}
\end{table}

Passing these tests confirms that the target variable exhibits chaotic properties suitable for modelling with the Chaos-Markov-Gaussian (CMG) framework. This validation is essential for justifying the use of nonlinear and probabilistic methods in subsequent modelling \cite{93,38}.

Across the studied indices, most exhibited positive Lyapunov exponents, confirming sensitive dependence on initial conditions \cite{95}. Correlation dimensions mostly ranged between 2.1 and 2.8, indicating low-dimensional chaos rather than pure randomness. Approximate and Sample Entropy values demonstrated irregularity and complexity, while consistently high spectral entropy reflected frequency-domain disorder. DFA exponents near 0.5 suggested a balance of short- and long-term dependencies. Collectively, these findings confirm chaotic dynamics across multiple indices and support the integration of chaos-based components in the CMG model.

These results collectively confirm that the target variable exhibits chaotic dynamics across multiple indices, justifying the integration of chaos-based modeling components in the proposed Chaos–Markov–Gaussian (CMG) framework.

\subsubsection{Temporal Standardization Strategy for Target and Feature Variables}

An important enhancement in the data preprocessing pipeline involves the integration of a day-wise, causally consistent standardization strategy, specifically designed for short-term sentiment forecasting tasks. Standardization is widely recognized for its critical benefits in deep learning, including stabilizing training behavior, accelerating convergence, preserving gradient stability, and improving performance on variable-length sequence inputs \cite{96,97}.

Given the sequential and time-sensitive nature of the problem—where the model aims to predict next-day sentiment—per-day normalization is applied to input features. For each trading day, features are z-score standardized using statistics (mean and variance) calculated only from that day’s data, ensuring no information is leaked from the future. This rolling window approach preserves temporal causality, allows the model to dynamically adjust to non-stationary input distributions, and supports evolving market dynamics and sentiment trends \cite{98,99}.

In contrast, the target variable, which represents the sentiment intensity in discrete classes, undergoes global normalization using all historical values up to the current observation. This global standardization supports stability and consistency across categorical forecasting targets and helps mitigate class imbalance and volatility effects \cite{100}.

This dual-path normalization framework—daily localized standardization for inputs and historical cumulative standardization for targets—preserves the model's temporal integrity, enhances generalization, and improves robustness against distributional shifts over time \cite{101,102}.

\begin{figure}[htbp]
    \centering
    \includegraphics[height=0.6\textwidth]{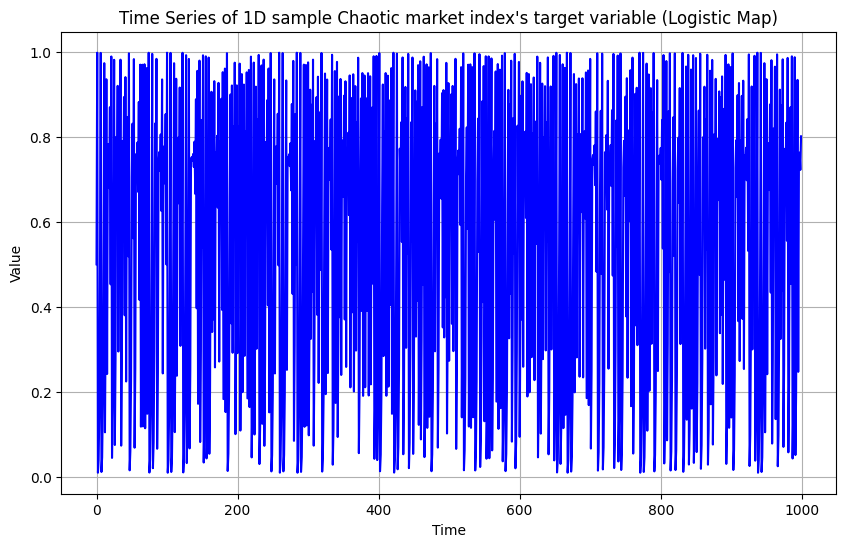}
    \caption{Time Series of a sample market index's target Variable}
    \label{fig:cmg_model}
\end{figure}
\begin{figure}[htbp]
    \centering
    \includegraphics[height=0.55\textwidth]{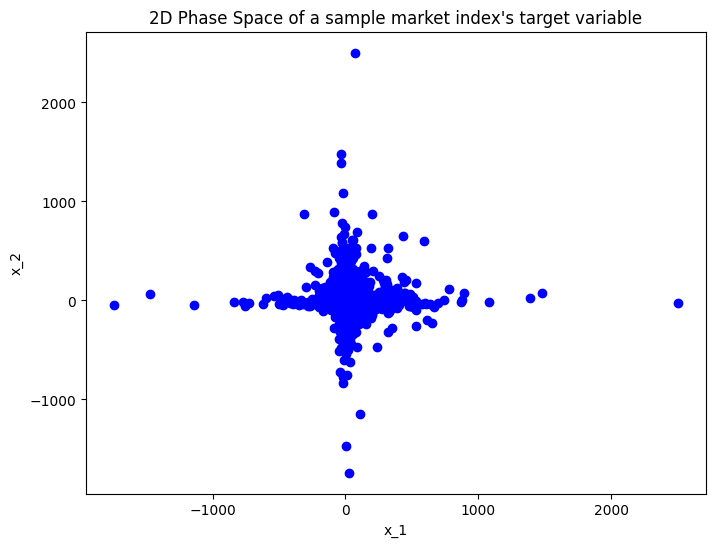}
    \caption{2D phase space of a sample market index's target variable}
    \label{fig:cmg_model}
\end{figure}

\begin{figure}[htbp]
    \centering
    \includegraphics[height=0.55\textwidth]{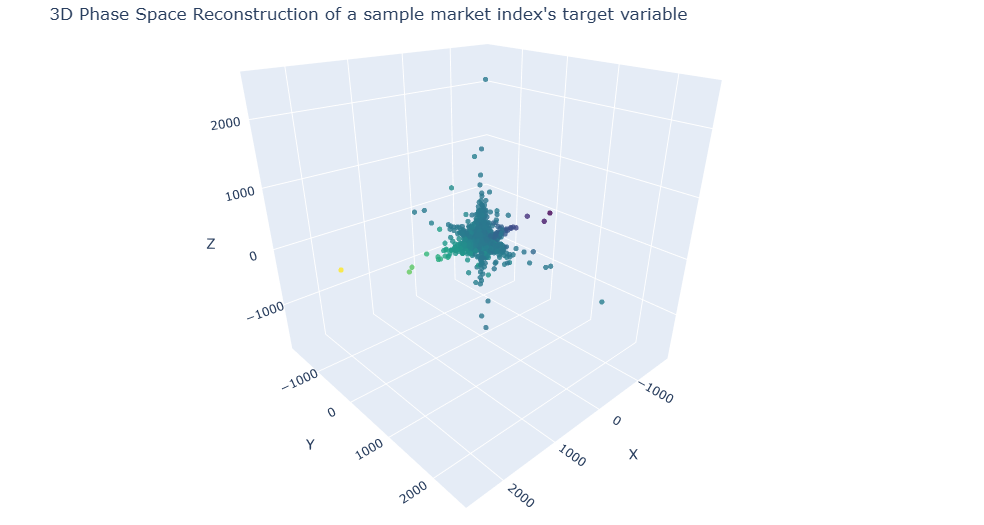}
    \caption{3D phase space of a sample market index's target variable}
    \label{fig:cmg_model}
\end{figure}

\begin{figure}[htbp]
    \centering
    \includegraphics[height=0.55\textwidth]{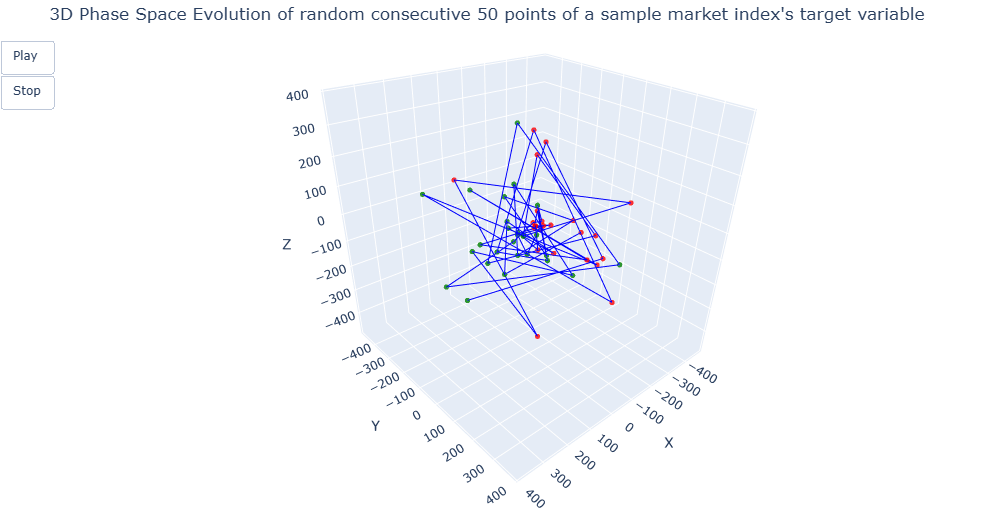}
    \caption{3D phase space Evolution of random consecutive 50 data points of a sample market index's target variable}
    \label{fig:cmg_model}
\end{figure}

\medskip
\medskip
\medskip
\medskip
\medskip
\medskip
\medskip
\medskip
\medskip
\medskip
\medskip
\medskip
\medskip
\medskip
\medskip
\medskip
\medskip
\medskip
\medskip
\medskip
\medskip
\medskip
\medskip
\subsection{Baseline Models}

A diverse set of baseline models is used to establish performance benchmarks across traditional, machine learning, and deep learning categories:
\subsubsection{Traditional Statistical \& Machine Learning Models:}

\begin{itemize}[noitemsep, topsep=0pt]
    \item Logistic Regression \cite{103}
    \item Gaussian Naive Bayes \cite{104}
    \item Multi Logistic Regression \cite{105}
    \item Random Forest Classifier \cite{106}
    \item XGBoost \cite{107}
\end{itemize}
\subsubsection{Deep Learning Models:}
\begin{itemize}[noitemsep, topsep=0pt]
    \item Long Short-Term Memory (LSTM) \cite{108}
    \item Gated Recurrent Unit (GRU) \cite{109}
    \item BiLSTM \cite{110}
\end{itemize}
\subsubsection{Proposed Model:}
\begin{itemize}[noitemsep, topsep=0pt]
    \item Chaos–Markov–Gaussian [CMG] 
\end{itemize}
\medskip

Each model was trained on the same set of input features derived from OHLC data and evaluated using identical target variables constructed from MACD signal intersections.

\section{Results}

To evaluate the effectiveness of the proposed Chaos-Markov-Gaussian (CMG) framework, we compared its predictive accuracy with a range of baseline models across 160 market indices from over 16 stock markets. The models were assessed based on their ability to predict daybreak sentiment, as described in Section 1. The average prediction accuracy across all indices is presented in Table-2.

\begin{table}[htbp]
\centering
\caption{Average Daybreak Sentiment Prediction Accuracy Across Models for 160 market indices}
\medskip
\renewcommand{\arraystretch}{1.2}  
\begin{tabular}{|l|c|}
\hline
\textbf{Model} & \textbf{Average Accuracy} \\
\hline
CMG & 0.7672 \\
LSTM & 0.7605 \\
BiLSTM & 0.7602 \\
GRU & 0.7601 \\
Logistic Regression & 0.7562 \\
XGBoost & 0.7550 \\
Random Forest Classifier & 0.7539 \\
Gaussian Naive Bayes & 0.7536 \\
Multi Logistic Regression & 0.7499 \\
\hline
\end{tabular}
\end{table}

As shown in the table, the CMG framework achieved the highest average accuracy (0.7672), outperforming all other models under identical experimental settings. Although the differences in accuracy values appear marginal, a statistical analysis was necessary to determine whether these improvements are statistically significant given the tightly controlled environment.

\subsection{Paired t-Test Results}

A paired t-test was conducted to compare the prediction accuracies of CMG with each baseline model across all individual market indices. This test assumes that the differences in prediction accuracy are normally distributed and is appropriate for comparing the means of two related groups \cite{111}. The p-values resulting from these tests are reported in Table 3.

\begin{table}[htbp]
\centering
\caption{Paired t-Test Results (CMG vs. Other Models)}
\medskip
\renewcommand{\arraystretch}{1.2}  
\begin{tabular}{|l|c|}
\hline
\textbf{Comparison} & \textbf{p-value} \\
\hline
CMG vs. BiLSTM & 0.0063 \\
CMG vs. Gaussian Naive Bayes & 0.0001 \\
CMG vs. GRU & 0.0153 \\
CMG vs. Logistic Regression & 0.0006 \\
CMG vs. LSTM & 0.0073 \\
CMG vs. Multi Logistic Reg. & 0.0000 \\
CMG vs. Random Forest & 0.0001 \\
CMG vs. XGBoost & 0.0008 \\
\hline
\end{tabular}
\end{table}

\subsection{Wilcoxon Signed-Rank Test Results}

To validate the robustness of these results without assuming normality, we also performed the Wilcoxon signed-rank test—a non-parametric alternative to the paired t-test that evaluates the median of the differences \cite{112}. The p-values from this analysis are shown in Table 4.

\begin{table}[htbp]
\centering
\caption{Wilcoxon Signed-Rank Test Results (CMG vs. Other Models)}
\medskip
\renewcommand{\arraystretch}{1.2}  
\begin{tabular}{|l|c|}
\hline
\textbf{Comparison} & \textbf{p-value} \\
\hline
CMG vs. BiLSTM & 0.00399 \\
CMG vs. Gaussian Naive Bayes & 0.00012 \\
CMG vs. GRU & 0.02520 \\
CMG vs. Logistic Regression & 0.00106 \\
CMG vs. LSTM & 0.01004 \\
CMG vs. Multi Logistic Reg. & $2.93 \times 10^{-7}$ \\
CMG vs. Random Forest & 0.00027 \\
CMG vs. XGBoost & 0.00160 \\
\hline
\end{tabular}
\end{table}

\subsection{Summary and Interpretation}

Both the paired t-test and the Wilcoxon signed-rank test consistently show that the CMG framework achieves statistically significant improvements in predictive performance over all other models. In each case, the p-value is below the standard 0.05 significance level, providing strong evidence against the null hypothesis that the CMG framework's accuracy is equivalent to that of the competing models.
\medskip

The statistical tests confirm that the observed improvements are not merely due to random chance but reflect a genuine performance enhancement attributable to the CMG framework. Despite the small magnitude of difference in average accuracy—which is expected in a rigorously constrained and standardized experimental setting—the consistent statistical significance across both tests demonstrates the value addition of the CMG model. This finding supports the conclusion that the CMG framework introduces meaningful innovations in sentiment prediction, particularly in high-volatility, event-driven time frames like market openings.
\medskip

In summary, the experimental results, reinforced by robust statistical validation, affirm the superiority of the CMG framework over traditional machine learning and deep learning models in the context of financial sentiment prediction.

\section{Opportunities for Model Enhancement}

The current experiment was conducted in a constrained environment with several restrictions to accommodate limited resources. To improve the accuracy and generalizability of the CMG framework, it is essential to relax these constraints and consider enhancements suitable for generic instruments, such as OHLCV (Open, High, Low, Close, Volume) data.

\subsection{Incorporation of Additional Features:}

One immediate improvement is to include the Volume variable of the instrument, which often contains valuable information about market activity and liquidity that can enhance predictive performance \cite{113}.

\subsection{Model Complexity and Resources:}

The transformer architecture used in this study consisted of only a single layer. Increasing computational resources would allow the use of deeper models with higher complexity, including:

\begin{itemize}[noitemsep, topsep=0pt]
    \item More transformer layers
    \item Larger inner dimensionality of the model (d\_model)
    \item A greater number of attention heads
    \item Increasing the number of LSTM cells in the final layer
\end{itemize}
Such enhancements can significantly improve the model’s capacity to learn complex patterns and dependencies \cite{114}.
\medskip
\medskip
\medskip
\medskip

\begin{table}[h]
\centering
\caption{Model Size Comparison Used in the Experiment}
\medskip
\begin{tabular}{|l|c|}
\hline
\textbf{Model} & \textbf{Size (KB)} \\
\hline
CMG     & 2194  \\
BiLSTM  & 3557  \\
GRU     & 1354  \\
LSTM    & 1786  \\
\hline
\end{tabular}
\label{tab:model_size_comparison}
\end{table}

\subsection{Generalization and Market Coverage:}

Transformers excel in generalization and can leverage large-scale training data. By scaling resources exponentially, it is possible to train a single transformer model on all market indices, thus capturing broad market dynamics. Alternatively, creating market-specific transformers trained on batches of highly correlated market indices can exploit inter-market relationships to further improve sentiment prediction accuracy \cite{115}.

\subsection{Data Preprocessing and Forecast Horizon:}

Altering the data preprocessing and data standardization steps to extend the forecast horizon from short-term to mid- or long-term predictions is another promising direction. However, extending the timeframe requires more than technical analysis alone. Variables related to fundamental analysis—such as macroeconomic indicators and news sentiment—become increasingly important. Transformers’ proven ability to process textual context efficiently makes them ideal for integrating such diverse data sources \cite{116}.

A potential approach is to employ ensemble models that separately process technical and fundamental inputs, for example, using architectures with multiple encoders (e.g., two encoders for technical and fundamental data) feeding into a common decoder. This allows the model to learn from both data domains effectively.

\subsection{Dynamic Attention Mechanism:}
Incorporating a dynamic attention mechanism where attention masks are used to prioritize states or tokens that significantly influence the target variable can enhance training efficiency and model interpretability. This focused learning can improve the model’s ability to capture meaningful temporal patterns \cite{117}.
\subsection{Multi-Horizon Prediction:}

Implementing multi-horizon prediction, where the model simultaneously forecasts multiple future steps (e.g., t+1, t+2, t+3), requires expanding the CMG framework to include n distinct output heads, one for each prediction horizon, and augmenting the target variables accordingly. This approach enables the model to learn generalized temporal dynamics across varying forecast intervals, helping to reduce overfitting to short-term noise and improve long-term robustness \cite{118}.

To effectively implement this within the CMG architecture, we adopt a 1 encoder – n decoder configuration. Here, a single shared encoder processes the input sequence and extracts a unified latent representation of the time series. Each decoder is then specialized to learn the temporal dependencies and nonlinear transformations associated with a specific prediction horizon. This design not only promotes parameter sharing and training efficiency but also ensures that each decoder focuses on horizon-specific error correction, enabling more accurate and stable multi-step forecasting.

\subsection{Cross-Market Transfer Learning:}

Incorporating data from multiple related markets can enhance sentiment prediction accuracy. For example, when forecasting sentiments for Asian markets, integrating recent immediate OHLCV data from both Asian and US markets simultaneously allows the model to capture cross-market influences and interdependencies. Including such multi-market variables across different time horizons enables a more comprehensive understanding of global market dynamics, thereby improving the robustness and precision of short-term sentiment forecasts \cite{119}.

\section{Conclusion}
This paper presented the Chaos-Markov-Gaussian (CMG) framework, a novel approach to short-term market sentiment prediction that integrates chaos theory, Markov properties, and Gaussian modelling within a transformer-based architecture enhanced by LSTM layers. The framework was rigorously evaluated on a comprehensive dataset spanning 160 market indices across more than 16 global stock markets.

The experimental results demonstrate that the CMG framework consistently outperforms a broad range of established baseline models, including traditional machine learning classifiers and deep learning architectures such as BiLSTM, GRU, and LSTM. Although the improvement in average prediction accuracy appears incremental, statistical analyses using paired t-tests and Wilcoxon signed-rank tests confirm that these enhancements are significant and not attributable to random variation.

By focusing on event-driven timeframes such as daybreak sentiment prediction, the CMG framework offers a robust and innovative methodology for capturing the complex dynamics inherent in financial time series data. The one-to-one model-to-market index approach, combined with the incorporation of CMG principles, allows the framework to operate effectively in resource-constrained environments while maintaining high predictive performance.

Future work can build on these findings by expanding the framework to multi-horizon predictions, integrating fundamental market signals, and exploring ensemble architectures that leverage both technical and textual data sources. Overall, the CMG framework establishes a promising direction for advancing the accuracy and applicability of sentiment prediction models in financial markets.

\medskip

\appendix

\section{Appendix A: List of Market Indices Used in the Study}
\label{appendix:A}
The following 160 market indices, identified by their Yahoo Finance tickers, were used as part of the experimental dataset for short-term sentiment modelling in the CMG framework:

\begin{quote}
\texttt{[CNXFMCG, XBANK, NIFTYQUALITY30, DJUSRD, DJUSMC, NIFTY100LOWVOL30, BSE.SMEIPO, TRMI, TFBI, BSE.PR.OWER, DJUSDN, LRGCAP, BSE.500, NIFTYALPHALOWVOL, CNXCONSUM, CNX100, LMI250, DJUSTK, TSSI, TBNI, DFMGI, NIFTY100ESGSECLDR, DJUSIP, DJUSDR, NIFTYMIDLIQ15, NLIN, QERP, MXX, NSEI, INFRASTRUCTURE, BSEMOI, NIFTY\_MIDCAP\_100, DJUSTB, CNXPHARMA, BSE.100, ESG100, CNXREALTY, CNXINFRA, TMDI, TTNI, DJUSIQ, DJUSPB, DJUSWU, TELCOM, NIFTY200QUALTY30, NIFTY50EQUALWEIGHT, QEAS, QBNK, DJUSRR, XOI, DJUSAI, CNXMETAL, BSE.MIDCAP, DFRGRI, DJUSLE, DJUSAR, DJUSOI, DJUSOS, QETR, DJUSGI, CNX200, DJUSSD, KQ184, CNXCMDT, BSE.BANK, CNXPSUBANK, BHRT22, TUTI, NIFTY200MOMENTM30, NIFTY100\_EQL\_WGT, BSEQUI, BSEDSI, MFG, NLHC, DJUSST, INDN, TASI, BSE.METAL, BSE.CG, TDAI, DJUSOQ, NSEBANK, DJUSRS, GSPTTCS, GSPE, FTSE, DOLLEX30, DJUSVN, TISI, DJUSIM, DJUSME, NSEMDCP50, DJUSIT, NIFTYSML\\CAP50, DJUSHV, DJUSPL, CNXPSE, DJUSUO, CNXMEDIA, MIDSEL, CNXIT, DJUSCX, DJUSAF, QIND, GSPC, CNXFIN, TXIE, DJT, DJUSNF, BSEEVI, DJUSSP, SP400, QERI, NIFTYMIDSML\\400, NI15, TX60, BSE.AUTO, NIFTY\_PVT\_BANK, DJUSAU, DJUSHG, CRSMID, DJI2MN, DJUSRT, DJUSFP, DJUSCC, DJUSCN, DJUSCP, DJUSNS, DJUSRH, SPTMI, DJUSPC, SP1500, BSE.PR.SU, DJUSIL, DJUSHR, NIFTYMIDCAP150, BSESN, TTSI, DJUSNG, INDSTR, DJUSIN, BSE.IPO, BEOGP, SPCBMIRWDUSD1, KRX100EW, DJUSBE, DJUSNC, BSE.TECK, DJUSEN, DJUSRU, DJUSOL, CNXENERGY, DJUSFE, BSE.REALTY, BELSC, BELMG, DJUSAS, DJUSEE, AMX, EZ300]}
\end{quote}

\section{Appendix B: List of Variable Names (Columns/Independent Variables)}
\label{appendix:B}

A full list of variable names (columns/independent variables) used as indicators in this study can be found below. This sample illustrates the format:

\begin{quote}
[Open, High, Low, Close, ABER\_ZG\_5\_15, ABER\_SG\_5\_15, ABER\_XG\_5\_15, ABER\_\\ATR\_5\_15, ACCBL\_20, ACCBM\_20, ACCBU\_20, ADX\_14, DMP\_14, DMN\_14, ALMA\_\\10\_6.0\_0.85, AMATe\_LR\_8\_21\_2, AMATe\_SR\_8\_21\_2, AO\_5\_34, APO\_12\_26, AROOND\_14, AROONU\_14, AROONOSC\_14, ATRr\_14, BBL\_5\_2.0, BBM\_5\_2.0, BBU\_5\_2.0, BBB\_5\_2.0, BBP\_5\_2.0, BIAS\_SMA\_26, BOP, AR\_26, BR\_26, CCI\_14\_0.015, CFO\_9, CG\_10, CHOP\_\\14\_1\_100, CKSPl\_10\_3\_20, CKSPs\_10\_3\_20, CMO\_14, COPC\_11\_14\_10, CTI\_12, LDECAY\_5, DEC\_1, DEMA\_10, DCL\_20\_20, DCM\_20\_20, DCU\_20\_20, EBSW\_40\_10, EMA\_10, ENTP\_10, ER\_10, BULLP\_13, BEARP\_13, FISHERT\_9\_1, FISHERTs\_9\_1, FWMA\_10, HA\_open, HA\_high, HA\_low, HA\_close, HL2, HLC3, HMA\_10, INC\_1, INERTIA\_20\_14, JMA\_7\_0, KCLe\_20\_2, KCBe\_20\_2, KCUe\_20\_2, K\_9\_3, D\_9\_3, J\_9\_3, KST\_10\_15\_20\_30\_10\_\\10\_10\_15, KSTs\_9, KURT\_30, LR\_14, LOGRET\_1, MACD\_12\_26\_9, MACDh\_12\_26\_9, MACDs\_12\_26\_9, MAD\_30, MASSI\_9\_25, MCGD\_10, MEDIAN\_30, MIDPOINT\_2, MIDPRICE\_2, MOM\_10, NATR\_14, OHLC4, PDIST, PCTRET\_1, PGO\_14, PPO\_12\_26\_9, PPOh\_12\_26\_9, PPOs\_12\_26\_9, PSARaf\_0.02\_0.2, PSARr\_0.02\_0.2, PWMA\_10, QS\_10, QTL\_30\_0.5, RMA\_10, ROC\_10, RSI\_14, RSX\_14, RVGI\_14\_4, RVGIs\_14\_4, RVI\_14, SINWMA\_14, SKEW\_30, SLOPE\_1, SMA\_10, SMI\_5\_20\_5, SMIs\_5\_20\_5, SMIo\_5\_20\_5, SQZ\_20\_2.0\_20\_1.5, SQZ\_ON, SQZ\_OFF, SQZ\_NO, SQZPRO\_20\_2.0\_20\_2\_1.5\_1, SQZPRO\_\\ON\_WIDE, SQZPRO\_ON\_NORMAL, SQZPRO\_ON\_NARROW, SQZPRO\_OFF, SQZPRO\_\\NO, SSF\_10\_2, STC\_10\_12\_26\_0.5, STCmacd\_10\_12\_26\_0.5, STCstoch\_10\_12\_26\_0.5, STDEV\_\\30, STOCHk\_14\_3\_3, STOCHd\_14\_3\_3, STOCHRSIk\_14\_14\_3\_3, STOCHRSId\_14\_14\_3\_3, SUPERT\_7\_3.0, SUPERTd\_7\_3.0, SWMA\_10, T3\_10\_0.7, TEMA\_10, THERMO\_20\_2\_0.5, THERMOma\_20\_2\_0.5, THERMOl\_20\_2\_0.5, THERMOs\_20\_2\_0.5, TRIMA\_10, TRIX\_30\_9, TRIXs\_30\_9, TRUERANGE\_1, TTM\_TRND\_6, UI\_14, UO\_7\_14\_28, VAR\_30, VHF\_28, VTXP\_14, VTXM\_14, WCP, WILLR\_14, WMA\_10, ZL\_EMA\_10, CDL\_DOJI\_10\_0.1, CDL\_INSIDE, PSARsl, SUPERTsl,  time]
\end{quote}

\subsection*{B.1 Technical Indicator Generation}
The technical indicators were generated using the \texttt{pandas\_ta} library, a powerful technical analysis extension built on top of \texttt{pandas}, which enables seamless computation and integration of technical indicators into financial time series datasets.

\begin{itemize}
    \item \textbf{Library Used:} \texttt{pandas\_ta}
    \item \textbf{Python 3 Pandas Extension:} \url{https://pypi.org/project/pandas-ta/}
    \item \textbf{Purpose:} To compute a wide range of technical indicators for feature engineering in financial forecasting models.
\end{itemize}

\section*{Appendix C: Code Availability}
The complete source code for the CMG framework and baseline models used in this study is available at:

\begin{quote}
\url{https://github.com/Infi-flux/CMG} 
\end{quote}

This repository includes all necessary scripts for data preprocessing, model training, evaluation, and result replication.

\end{document}